\documentstyle[12pt,psfig,aasms4,flushrt]{article}

\makeatletter
\@ifundefined{chapter}{\def\thebibliography#1{\section*{References\@mkboth
  {REFERENCES}{REFERENCES}}\list
  {\relax}{\setlength{\labelsep}{0em}
        \setlength{\itemindent}{-\bibhang}
        \setlength{\itemsep}{0pt}
        \setlength{\parsep}{0pt}
        \setlength{\leftmargin}{\bibhang}}
    \def\newblock{\hskip .11em plus .33em minus .07em}
    \sloppy\clubpenalty4000\widowpenalty4000
    \sfcode`\.=1000\relax}}%
{\def\thebibliography#1{\chapter*{Bibliography\@mkboth
  {BIBLIOGRAPHY}{BIBLIOGRAPHY}}\list
  {\relax}{\setlength{\labelsep}{0em}
        \setlength{\itemindent}{-\bibhang}
        \setlength{\itemsep}{0pt}
        \setlength{\parsep}{0pt}
        \setlength{\leftmargin}{\bibhang}}
    \def\newblock{\hskip .11em plus .33em minus .07em}
    \sloppy\clubpenalty4000\widowpenalty4000
    \sfcode`\.=1000\relax}}

\newlength{\bibhang}
\let\@internalcite\cite
\def\cite{\let\@citeleft(\let\@citeright)%
    \@ifstar{\citeyear}{\citefull}}
\def\citenp{\let\@citeleft\relax\let\@citeright\relax
    \@ifstar{\citeyear}{\citefull}}
\def\citefull{\def\astroncite##1##2{##1~##2}\@internalcite}
\def\citeyear{\def\astroncite##1##2{##2}\@internalcite}

\def\@citex[#1]#2{\if@filesw\immediate\write\@auxout{\string\citation{#2}}\fi
  \def\@citea{}\@cite{\@for\@citeb:=#2\do
    {\@citea\def\@citea{; }\@ifundefined
       {b@\@citeb}{{\bf ?}\@warning
       {Citation `\@citeb' on page \thepage \space undefined}}%
{\csname b@\@citeb\endcsname}}}{#1}}

\def\@cite#1#2{\@citeleft#1\if@tempswa , #2\fi\@citeright}
\def\@biblabel#1{}

\makeatother

\newcommand{\squeeze}{\setlength{\baselineskip}{0.7\baselineskip}}
\newcommand{\PSbox}[2]{\centerline{~\psfig{file=#1,width=#2}~}}
\newcommand{\PSbound}[3]{\centerline{~\psfig{file=#1,width=#2,#3}~}}

\renewcommand\approx{\mbox{$\sim$}}

\newcommand\approxgt{\mbox{$^{>}\hspace{-0.24cm}_{\sim}$}}
\newcommand\approxlt{\mbox{$^{<}\hspace{-0.24cm}_{\sim}$}}

\newcommand\etal{et~al.$\!$}
\newcommand{\kmsec}{\mbox{km sec$^{-1}$}}
\newcommand{\SzX}{\mbox{$S_{0{\rm,8\,GHz}}$}}
\newcommand{\tzX}{\mbox{$t_{0{\rm,8\,GHz}}$}}

\newcommand{\HmG}[1]{\mbox{$H_{0}^{#1}$}}
\newcommand{\ppm}{\mbox{$\pm$}}
\newcommand\cf{{\it cf.}}
\def\arcsec{{$^{\prime\prime}$}}
\def\rxte{{{\it RXTE}}}
\def\rxtez{{{\it RXTE\/}}}

\def\einsteinz{{{\it Einstein\/}}}

\def\axafz{{{\it AXAF\/}}}
\def\ujy{{$\mu$Jy}}
\def\uJy{\ujy}
\def\fg{{FG}}
\newcommand{\tee}[1]{\mbox{$\times 10^{#1}$}}
\def\cgsflux{{erg~cm$^{-2}$~s$^{-1}$}}
\received{21 December 1998}
\accepted{10 November 1999}
\begin{document}

\squeeze

\title{Identification of a Likely Radio Counterpart\\ of the Rapid
  Burster} 

\author{Christopher B. Moore\altaffilmark{1}, Robert E.
  Rutledge\altaffilmark{2}, Derek W. Fox\altaffilmark{3}, Robert A.  Guerriero\altaffilmark{4}, Walter H.~G.  Lewin\altaffilmark{3}, Robert
  Fender\altaffilmark{5}, Jan van Paradijs\altaffilmark{5,6}}

\altaffiltext{1}{\squeeze Kapteyn Astronomical Institute, Postbus 800, 9700~AV Groningen,  The Netherlands; e-mail: {\it cmoore@cfa.harvard.edu}}
\altaffiltext{2}{\squeeze Space Radiation Laboratory, California Institute of Technology, 
MC 220-47, Pasadena, CA 91125; e-mail: {\it rutledge@srl.caltech.edu}}
\altaffiltext{3}{\squeeze Massachusetts Institute of Technology, Dept.~of
Physics, 37-627, Cambridge, MA 02139; e-mail: {\it
derekfox@space.mit.edu,
lewin@space.mit.edu}}
\altaffiltext{4}{\squeeze United States Military Academy, Department of
Physics, West Point, NY 10996; e-mail: {\it hr6958@exmail.usma.army.mil}
}
\altaffiltext{5}{\squeeze Astronomical Institute ``Anton Pannekoek'', University of
Amsterdam, Center for High Energy Astrophysics, Kruislaan 403, 1098 SJ
Amsterdam, The Netherlands; email: {\it
rpf@astro.uva.nl}}
\altaffiltext{6}{\squeeze University of Alabama, Huntsville; e-mail {\it jvp@astro.uva.nl }}

\begin{abstract}
  We have identified a likely radio counterpart to the low-mass X-ray
  binary MXB~1730$-$335 (the Rapid Burster).  The counterpart has
  shown 8.4~GHz radio on/off behavior correlated with the X-ray on/off
  behavior as observed by the \rxte/ASM during six VLA observations.
  The probability of an unrelated, randomly varying background source
  duplicating this behavior is 1--3\% depending on the correlation
  time scale.  The location of the radio source is RA~$17^{\rm
  h}33^{\rm m}24\fs61$; Dec~$-$33\arcdeg 23\arcmin 19\farcs 8 (J2000),
  \ppm 0\farcs1.  We do not detect 8.4~GHz radio emission coincident
  with type~II (accretion-driven) X-ray bursts.  The ratio of radio to
  X-ray emission during such bursts is constrained to be below the
  ratio observed during X-ray persistent emission at the 2.9$\sigma$
  level.  Synchrotron bubble models of the radio emission can provide
  a reasonable fit to the full data set, collected over several
  outbursts, assuming that the radio evolution is the same from
  outburst to outburst, but given the physical constraints the
  emission is more likely to be due to $\sim$hour-long radio flares
  such as have been observed from the X-ray binary GRS~1915+105.
\end{abstract}

\section{Introduction}

Discovered in 1976 \cite{lewin76}, the Rapid Burster (MXB 1730$-$335,
hereafter ``RB'') is located in the highly reddened globular cluster
Liller~1 \cite{liller77}, which has a distance modulus of
14.68\ppm0.23, corresponding to 8.6\ppm1.1 kpc, determined by
main-sequence fitting \cite{frogel95}.  Liller~1 has a small optical
core radius (about 6\farcs5).    

Radio observations of transient X-ray binaries have found several to
be radio transients as well (see \citenp{hjellming95} for a review),
both black hole candidates (A0620$-$00, GS~2000+25, GS~2023+33,
GS~1124$-$68) and neutron stars (Aql~X-1, Cir~X-1, Cen~X-4).  The
outbursts of X-ray transients are due to a sudden turn-on of accretion
onto the compact object in a binary lasting from $\sim$days to months
\cite{tanakalewin95,chen97}.  The radio spectral and temporal behavior
in some of these objects is described by a synchrotron bubble model
\cite{vanderlaan66} which indicates that there are plasma outflows
associated with the X-ray outburst.  In particular, the black hole
candidate and superluminal jet source GRS~1915+105 has been seen to
exhibit correlated behavior at X-ray, infrared and radio wavelengths
\cite{fender97,mirabel98,eik98a,eik98b,fender98} that is readily
understood in terms of the synchrotron emission of expanding plasmoids
that have been ejected from the inner regions of the system
\cite{fender97,mirabel98,fender98}.

Despite its frequent X-ray outbursts (average interval, $\sim220$
days; \citenp{bobg98}), several previous studies have not detected
radio emission from the RB
\cite{johnson78,lawrence83,grindlay86,johnston91,fruchter95}.
Improvements in radio sensitivity since some of those studies were
performed made the detection of a RB radio counterpart at previously
unattainable flux levels practical.  Furthermore, the advent of the
\rxtez\ All-Sky Monitor made it possible to correlate radio
observations with a well sampled X-ray light curve.  We therefore
undertook a search for a radio counterpart of the Rapid Burster in
X-ray outburst.  Since the type~II X-ray bursts of the Rapid Burster
are thought to be caused by the same phenomenon (spasmodic accretion)
as the X-ray outbursts \cite{jvp96}, one might expect that they are
accompanied by simultaneous radio emission.  The observations
reported here marginally exclude (at the 2.9$\sigma$ level)
simultaneous radio burst emission from the likely radio counterpart.

\subsection{X-ray behavior}

The RB is a transient X-ray source which has been observed during the
past few years to go into outburst approximately every 220 days for a
period of $\sim$30 days \cite{bobg98}.  It is the only low-mass X-ray
binary (LMXB) which produces two different types of X-ray bursts
\cite{hoffman78}.  Type~I bursts, which are observed from $\sim$40
other LMXBs, are due to thermonuclear flashes on the surface of an
accreting neutron star.  Type~II bursts, which have been observed
from only one other LMXB (GRO~J1744$-$28; \citenp{chryssa96,lewin96}),
are sudden releases of gravitational potential energy resulting from
accretion instabilities.  For a detailed review of type~I and type~II
bursts, see Lewin, van~Paradijs and Taam \cite*{lvt93,lewin95}.

The RB does not fit easily into the ``Z/Atoll'' paradigm of LMXBs
\cite{rutledge95b}, in which the correlated X-ray fast-timing and
spectral behavior of LMXBs divides the population into two classes
\cite{hvdk89,mvdk95}.  However, the RB is known to exhibit periods of
behavior characteristic of an Atoll source.  During an outburst in
1983, the RB exhibited strong persistent emission (PE) and type~I
bursts with no type~II bursts -- behavior typical of the Atoll sources
\cite{barr87}.  This behavior has also been seen in outbursts observed
more recently with \rxte.  For the first $\sim$17 days of these
outbursts, strong PE is accompanied only by type~I X-ray bursts.
Type~II bursting behavior begins after this period and continues while
the PE decreases (details of these observations are given in
\citenp{bobg98}).

In spite of 20 years of theoretical modeling, no satisfactory model
exists for the disk instability which drives the type~II bursts (for a
review, see \citenp{lewin95}).  Models which require weak magnetic
fields ($\approxlt 10^9$ G) were effectively excluded with the
discovery of GRO~J1744$-$28, which contains a pulsar with a magnetic
field of $\approx 10^{11}$ G \cite{finger96,cui97}, and also exhibits
type~II bursts \cite{chryssa96,lewin96}.

\subsection{Previous Radio Observations}

There have been several radio studies of fields of view containing
Liller~1, some with the goal of identifying a radio counterpart of the
Rapid Burster.  Some have focussed on finding radio variability in
hopes of catching bursts in the radio correlated with type~II X-ray
bursts while others have sought persistent radio sources.  We
summarize the most stringent limits on the flux density of a
persistent radio source in Table~\ref{tab:previous}; additional, less
stringent limits are summarized by Lawrence \etal\ \cite*{lawrence83}.
\begin{table}[tb]
\scriptsize
\begin{center}
\caption{Previous Radio Observations of Persistent Source\label{tab:previous} }
\vspace{2mm}
\begin{tabular}{lccccc} \tableline
{Reference}     & {Frequency}   &{Measurement}  & {Epoch}       & {Rapid Burster} & {Note}      \nl
{}              & {(GHz)}       &{(\uJy)}       & {}            & {State}         &                     \nl \tableline
\citenp{fruchter95}     & 0.33                  & 9000$\pm$1000         & 1-3 Aug 1990          & Unknown       & VLA DnC \nl
                        & 1.5                   & 280$\pm$ 50           & 25 May 1993           & Unknown       & VLA CnB \nl
                        & 4.5                   & 95$\pm$ 14            & Jul 1990-Jan 1991     & Unknown       & 11 Obs.; VLA C, B \nl
\citenp{johnston91}     & 1.5                   & $<$180 (4$\sigma$)    & 9-14 Apr 1990         & Unknown       & VLA A \nl
\citenp{grindlay86}     & 4.5                   & $<$380 (3$\sigma$)    & 21-22 Jun 1982        & Unknown       & VLA A \nl
\citenp{johnson78}      & 2.7                   & (3$\pm$2)\tee{3}      & 23, 24, 27 Apr 1977   & Bursting      & NRAO Greenbank \nl
                        & 8.1                   & (5$\pm$3)\tee{3}      & 23, 24, 27 Apr 1977   & Bursting      &                \nl
\citenp{lawrence83}     & 15.5                  & $<$93\tee{3} (3$\sigma$)  & 20 Apr 1980           & Not Bursting  & Haystack Obs. \nl
                        & 14.5                  & $<$294\tee{3} (3$\sigma$) &  17 Apr-28 Jul 1980   & Not Bursting  & U of Michigan \nl
                        &  8                    & $<$636\tee{3} (3$\sigma$) &  17 Apr-28 Jul 1980   & Not Bursting  &               \nl
                        &  5                    & $<$60\tee{3} (3$\sigma$)  & 31 Jul - 4 Aug 1980   & Not Bursting  & Parkes, CSIRO \nl
\tableline
\end{tabular}
\end{center}
{\small\squeeze Note.\ --- Additional observational limits on variable
radio emission may be found in Lawrence \etal\ (1983); periods when
the RB was ``not bursting'' may have had persistent emission, which
was difficult to distinguish from the nearby source MXB 1728$-$34.}
\end{table}

Fruchter \& Goss \cite*[hereafter, \fg]{fruchter95} 
discovered a radio source in three bands (0.33, 1.5, and 4.5GHz), with
a spectral slope of $\sim -2$.  They observe the flux density of this
source at 1.5~GHz above the upper limit derived from an observation at
another time \cite{johnston91}.  \fg\ interpret this discrepancy as
the result of beam dilution in the Johnston et al.\ observations
(0\farcs5 beam obtained with the VLA in A-array) over-resolving a
large population of radio pulsars in Liller~1.  Thus, the
interpretation of this radio source was as an integration over a
population of radio pulsars in Liller~1.  To date, no radio pulsations
have been detected from the direction of Liller~1 (R.  Manchester,
private communication).

Prior to the present work, there have been two studies which produced
limits on radio emission simultaneous with type~II X-ray bursts of the
RB.  Johnson \etal\ \cite*{johnson78} observed simultaneously in the
radio and X-ray bands during a period when a total of 64 type~II
bursts were observed in the X-ray, and placed upper limits on the
simultaneous flux density of radio bursts (at 2.7 and 8.1 GHz) of
$\sim$20~mJy.  Rao \& Venugopal \cite*{rao80} observed at 0.33 GHz
during two X-ray bursts and placed an upper limit of 0.2~Jy on
the radio flux density during the bursts. 

There are claims of radio burst detections from the RB without
simultaneous X-ray observations (\citenp{calla79,calla80a,calla80b};
and Calla, private communication in \citenp{johnson78}).  Calla \etal\
report observing approximately 14 radio bursts on nine different days
with peak flux densities of 400--600~Jy at 4.1~GHz and durations of
10--500~s.  This phenomenon has not, however, been confirmed at other
observatories and the reported flux densities are substantially above
the limits placed by Johnson \etal\ \cite*{johnson78} during their
simultaneous X-ray/radio observations.  If the reported radio bursts
are real, they do not appear to be correlated with type~II X-ray
bursts \cite{lawrence83}.  Using the 16.7~ksec of observations at
8.4~GHz obtained in the course of this work, we find a 3$\sigma$ upper
limit of 250~mJy on the flux density of our likely radio counterpart
during any single 3.3~sec integration.  Thus we see no evidence for
radio flares of the type reported by Calla \etal\ during either the
X-ray active or quiescent periods.

\subsection{Objectives}

The goals of the present work are, first, to search for a radio
counterpart of the Rapid Burster, detectable either in X-ray outburst
or in quiescence; and second, to determine if the radio emission can
be tied to the active accretion during an outburst. 

In Section~\ref{sec:observations} we describe the radio and X-ray
observations, and present some general results.  In
Section~\ref{sec:results}, we present the results of our analyses of the
radio observations in the context of the ``standard model'' of radio
emission from X-ray transients: the synchrotron bubble model.  We
evaluate the likelihood of the counterpart identification in
Section~\ref{sec:likelihood}, discuss the results of these observations
in Section~\ref{sec:discuss}, and list our conclusions in
Section~\ref{sec:conclusions}.

\section{Observations} \label{sec:observations}

X-ray observations were made with two instruments on the {\it Rossi}
X-ray Timing Explorer (\rxte; \citenp{bradt93}): the All-Sky Monitor
(ASM; \citenp{levine96}) and the Proportional Counter Array (PCA;
\citenp{zhang93,jahoda96}).  The ASM consists of three 1.5--12 keV
X-ray proportional counters with coded-aperture masks.  It obtains
observations of approximately 80\% of the sky every 90 minutes and is
sensitive to persistent X-ray sources down to $\sim$5~mCrab ($3\sigma$
detection) in a typical day's cumulative exposure \cite{remillard97}.
The standard data products of the ASM include the daily-average count
rate of a known catalogue of X-ray sources including the
RB\footnote{\small see
http://space.mit.edu/$\sim$derekfox/xte/ASM.html}.  A summary of our
radio observations and the contemporaneous ASM X-ray flux measurements
is shown in Table~\ref{tab:radio}.
\begin{table}[t!b]
\scriptsize
\begin{center}
\caption{Radio and X-ray Observations Results \label{tab:radio} }
\vspace{2mm}
\begin{tabular}{lccccc}\tableline
Date     & Observatory          &$\nu$       & $S_\nu$ & \rxte/ASM 1-day Avg & \rxte/PCA \\
(UT)     & (configuration)      &(GHz)       & (\uJy)  & (c/s)               & (c/s)     \\ \tableline
1996 Oct 14         &VLA (D$\rightarrow$A)         & 8.4                   & 45$\pm$30        &0.15$\pm$0.31                  & --                    \nl
1996 Nov 06.8       &VLA (A)                       & 8.4                   &370$\pm$ 45       &9.8$\pm$ 0.8                   & 2830$\pm$150          \nl
1996 Nov 11.9       &VLA (A)                       & 4.9                   &190$\pm$ 45       &10.2$\pm$ 1.5                  & 1660$\pm$100          \nl
''                 & ''                           & 8.4                   &310$\pm$ 35       & ''                            & ''                    \nl
1997 Jun 29.3      & VLA (C)                      & 4.9                   &210$\pm$ 70       &17.8$\pm$0.6                   & 3680$\pm$200          \nl
''                 & ''                           & 8.4                   &310$\pm$45        & ''                            & ''                    \nl
1997 Jul 24.1      & VLA (CS)                     & 8.4                   &41$\pm$30         &0.80$\pm$0.73                  & 210$\pm$100           \nl 
1998 Jan 30.8      &SCUBA                         & 350                   &$<$3000           & 20.4$\pm$0.4                  & --                    \nl
1998 Jan 31.1      &ATCA (6A)                     & 4.8                   &$<$390            & 16.6$\pm$0.9                  & --                    \nl
''                 & ''                           & 8.6                   &$<$360            & ''                            & --                    \nl
1998 Feb 08.1      & ATCA (6A)                    & 4.8                   &$<$480            & 6.88\ppm 0.31                 & --                    \nl
''                 & ''                           & 8.6                   &$<$930            & ''                            & --                    \nl
1998 Feb 19.6      &VLA (D$\rightarrow$A)         & 8.4                   &13\ppm30          & 2.72\ppm0.35                  & 360\ppm 120             \nl \hline
\multicolumn{5}{c}{Simultaneous Radio/X-ray Observations of Type~II Bursts} \nl
B1997 Jul 24.1     & VLA (CS)                     & 8.4                   &$-300\pm 230$     & --                            & 3240\ppm 100           \nl 
B1998 Feb 19.6\tablenotemark{a}  & VLA (D$\rightarrow$A)        & 8.4                   & 42\ppm 120       & --                           & 3550\ppm 100        \nl
B1998 Feb 19.6\tablenotemark{b}  & VLA (D$\rightarrow$A)        & 8.4                   & 74 \ppm 82        & --                            & 2350\ppm 100          \nl\tableline
\end{tabular}
\end{center}
{\small $^{\it a}$Using 3.33 sec bins with $>$
  1700 c/s average PCA countrate.} \\
{\small $^{\it b}$Using 3.33 sec bins with $>$ 300
  c/s average PCA countrate.} \\
{\small Note.\ --- Upper limits are 3$\sigma$}
\end{table}
The PCA is a collimated array of gas-filled proportional counters,
sensitive in the 2--60~keV energy range, with a FOV of $\sim 1^\circ$
and a geometric area of $\sim 6500\, {\rm cm}^2$.  In total, five
simultaneous PCA and radio observations were performed; they are
discussed in detail below.  When the PCA is pointed directly at the
RB, there is a nearby ($\sim0.5^\circ$) persistent X-ray source (4U
1728$-$34) which contributes to the measured X-ray flux.  During part
of these observations, the PCA is pointed offset from the RB at a
reduced collimator efficiency, excluding 4U~1728$-$34 entirely from
the field of view.  The uncertainty in the aspect correction of the
PCA dominates our uncertainty in the RB count rates, which throughout
this paper are given as the count rate for the RB only (aspect
corrected, background subtracted, with the count rate from
4U~1728$-$34 -- assumed constant over the 1-hr observation --
subtracted).

The radio observations were performed at three different
observatories.  The Very Large Array (VLA)\footnote{\small\squeeze The
VLA is part of the National Radio Astronomy Observatory, which is
operated by Associated Universities,~Inc., under cooperative agreement
with the National Science Foundation.} observed every outburst
discussed in this work and made the first detection of the radio
counterpart reported here.  VLA observations were obtained in two
closely spaced bands, each of 50~{MHz} nominal bandwidth.  Each band
was observed in both right-circular and left-circular polarization
making the total observed bandwidth 100~{MHz} for each polarization.
The complex antenna gains were set using observations of a nearby
compact radio source (1744$-$312) and the flux density scale was
determined using observations of 3C~286.

The Rapid Burster was observed on two occasions with the Australia
Telescope Compact Array (ATCA)\footnote{\small\squeeze The Australia
Telescope is funded by the Commonwealth of Australia for operation as
a National Facility managed by the CSIRO.}.  The complex antenna gains
were set using observations of the nearby phase calibrator, PKS
B1657$-$261.  The observations were made simultaneously in two
orthogonal linear polarizations at 4800 and 8640~MHz with a bandwidth
of 128~MHz at each frequency. For both observations the array was in
the extended 6A configuration with baselines in the range 337--5939~m.

The Sub-millimeter Common User Bolometer Array (SCUBA;
\citenp{cunningham94}) on the James Clerk Maxwell Telescope
(JCMT)\footnote{\small\squeeze The James Clerk Maxwell Telescope is
operated by The Joint Astronomy Centre on behalf of the Particle
Physics and Astronomy Research Council of the United Kingdom, the
Netherlands Organisation for Scientific Research, and the National
Research Council of Canada.}  observed on 30~January~1998~UT using the
850~$\mu$m system in photometry mode.

To date, there have been five outbursts of the RB observed with
\rxte/ASM.  The progression of four of these outbursts, including the
evolution of the RB's bursting behavior, is described elsewhere
\cite{bobg98}.  During three of these outbursts, we performed radio
observations with the VLA, ATCA, or SCUBA, the results of which are
listed in Table~\ref{tab:radio}.  We describe the observations in
detail in the following sections.

\subsection{November 1996 Outburst}

Radio observations with the VLA in A-configuration were made at
8.4~GHz while the RB was in X-ray quiescence on 1996~October~14.
These produced low upper limits in both the radio (45\ppm 30 \uJy) and
X-ray (0.15\ppm 0.31 ASM c/s) bands (errors are $1\sigma$; 73~ASM c/s
$=$ 1~Crab).  

\begin{figure}[t!b]
\PSbox{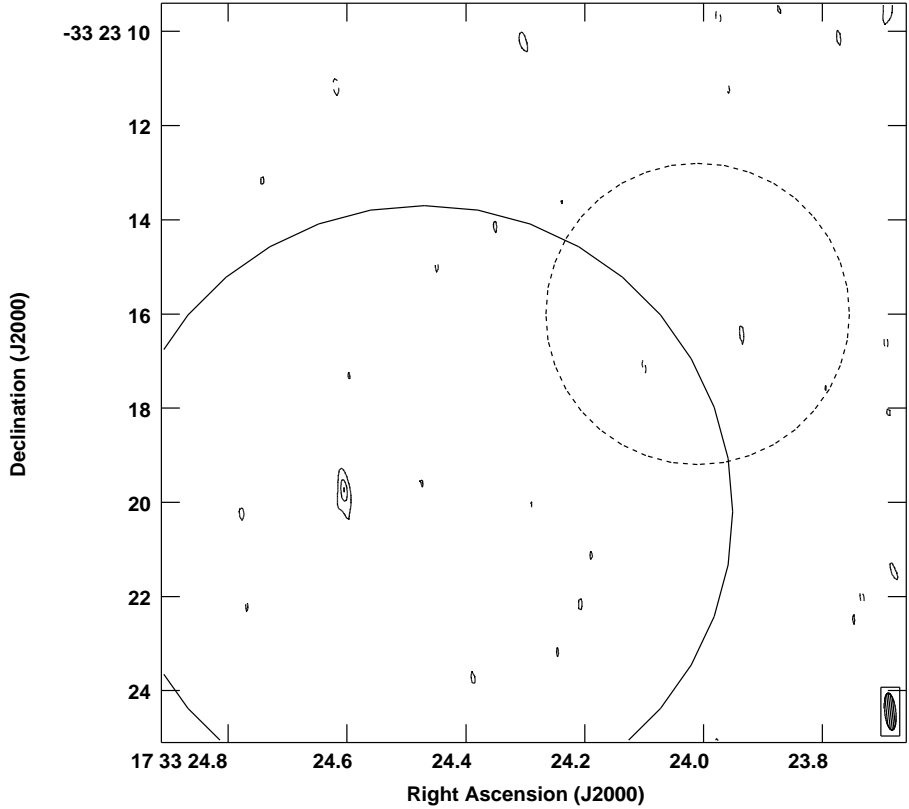}{12cm}
\caption{ \small \squeeze \label{fig:rbradio} {Discovery image}.  VLA
map at 8.4$\,$GHz of Liller 1 on 1996~Nov~6.8.  The size of the clean
beam is illustrated in the lower right corner.  The radio contours are
at 3$\sigma$, 6$\sigma$ and peak flux density.  The solid circle is
the 6\farcs5\ppm0\farcs5 core radius at the optical position for
Liller~1 (Kleinmann, Kleinmann \& Wright 1976, confirmed by Picard \&
Johnston 1995; which we adopt over the 3\farcs5\ppm0\farcs5 found by
Grindlay \etal\ 1984), which has absolute optical astrometric
uncertainty of 1\arcsec .  The dashed circle is the $2\sigma$ error
circle of the X-ray localization of the Rapid Burster (Grindlay \etal\
1984).}
\end{figure}

On 1996~October~29, the RB was detected by the ASM to have begun an
X-ray outburst, reaching peak intensity at 1996~October~30~14:30~UT
(\ppm 1 hr).  VLA observations, with simultaneous \rxte/PCA
observations took place 7.2~days after the time of X-ray peak flux
(1996~November~6.8~UT), at which time a radio source which was not
present in the October~14 observation was detected with a flux density
of 370\ppm45 \uJy (8.4 GHz). The new source was consistent with a
point source but the low signal-to-noise ratio limits the upper limit
on the size to 0\farcs5 in both dimensions.  The source is located
at RA~17h33m24s.61; Dec~$-$33d23m19s.8 (J2000), \ppm 0\farcs1 (see
Figure~\ref{fig:rbradio}; the positions of radio source detections
described below are consistent with this position).  
\nocite{kleinmann76,picard95,grindlay84}
During the 3.6~ksec PCA observation, one type~I X-ray burst was
observed along with strong persistent emission (PE; 220\ppm 10 mCrab;
throughout this work, 1 Crab=13,000 PCA c/s) but there were no type~II
X-ray bursts.  For the RB, we find an approximate conversion between
PCA count rate and X-ray flux to be 3\tee{-12} \cgsflux\ per PCA cps
(2-20 keV).

A second simultaneous VLA/PCA observation took place 12.3 days after
the X-ray maximum of the outburst (on 96~November~11.88~UT) at 4.89
and 8.44~GHz; the radio point source had flux densities of 190\ppm45
and 310\ppm35~\uJy (respectively).  During this observation, one
type~I X-ray burst was observed with persistent emission of
128\ppm8~mCrab and no type~II X-ray bursts.

\subsection{June-July 1997 Outburst}

The RB began the following X-ray outburst on 1997~June~25, reaching a
peak flux at 1997~June~26~10:30~UT (\ppm 2 hr).  Simultaneous VLA/PCA
observations took place 2.9~days after X-ray maximum while the VLA was in
C-array.  The radio object was present at both 4.89~GHz and 8.44~GHz,
with flux densities of 210\ppm70 and 310\ppm45~\uJy\ respectively.
During this observation, the PE observed by the PCA was
280\ppm15~mCrab and there were two type~I X-ray bursts but no type~II
X-ray bursts.

On 1997~July~24.08~UT, a second simultaneous VLA/PCA observation took
place, 27.6~days after X-ray maximum.  The VLA was in CS-array and the
radio point source was not detected at 41\ppm30~\uJy\ (8.44~GHz).  The
PE was very weak ($<$8~mCrab); over the whole observation, the average
intensity (bursts+PE) was 210\ppm100 PCA c/s (16~mCrab).  A total of
seven type~II X-ray bursts were observed with the PCA with an average
intensity of 2350\ppm100~PCA~c/s (180~mCrab) during the bursts, and a
mean duration of 12 seconds.  Of these, six had simultaneous coverage
at VLA at 8.44~GHz.  The radio data corresponding to these six bursts
were extracted to determine if there is detectable radio emission
during the brightest bursts (3240\ppm100 PCA c/s).  We find an upper
limit on the radio emission during these type~II X-ray bursts of
$<$690 \uJy ($3\sigma$).

\subsection{January-February 1998 Outburst}\label{sec:Jan98}

The next RB outburst began on 1998~January~27, peaking in X-ray
intensity at 1998~January~29~12:30~UT (\ppm 4~hours). During PCA
observations 1.2~days after X-ray maximum
(1998~January~30~19:19--21:09~UT), strong PE (4000\ppm100 c/s) was
observed along with two type~I X-ray bursts but no type~II X-ray
bursts.  Observations at the JCMT were carried out 1.3~days after the
X-ray maximum using the SCUBA system at 350~GHz and failed to detect
the proposed radio counterpart with a 3$\sigma$ upper limit of 3~mJy.

Beginning 1.6~days after the X-ray maximum, the ATCA observed for a
3-hour period (1:37-4:46~UT, on 1998~January~31) at 4.8 and
8.64~GHz. These observations produced flux density upper limits of
$<$390 and $<$360~\uJy\ respectively (3$\sigma$).  Between
1:33-4:28~UT on 1998~February~8, 9.6 days after the X-ray maximum,
observations at the ATCA produced $3\sigma$ upper limits of
$<$480~\uJy\ (4.80 GHz) and $<$930~\uJy\ (8.6 GHz).

On 1998~February~19, 20.1~days after the X-ray maximum, we performed a
7.8~ksec PCA observation during which a total of 91 type~II bursts
were observed.  Simultaneous VLA observations occurred during the
second half of this period and 39 type~II bursts were observed
simultaneously by the PCA and the VLA at 8.44~GHz.  These bursts had
average peak count rates of $\sim$6000~c/s (due only to type~II burst
emission, excluding background and PE of 250~PCA~c/s) and durations of $\sim$10
seconds.  Integrated over the full radio observation, the 3$\sigma$
upper limit on radio emission from a point source at the position of
the proposed radio counterpart was $<$90 \uJy\ (8.4 GHz).  This
observation is the only radio observation in the present work during
which a large number of type~II X-ray bursts were observed.
Therefore, we can use it to explore the relationship between the
type~II X-ray bursts and the radio emission.

In order to compare the X-ray flux to the radio flux density, we
re-bin the PCA data to correspond to the 3.33~s integration periods
of the VLA data.  To optimize the signal-to-noise ratio for the
detection of radio bursts under the assumption that the radio flux
density is proportional to the X-ray flux, we use only time bins with
average count rates $>$1700~c/s.  During these time bins, the 
radio emission is constrained to be $<$360~\uJy\ ($3\sigma$) at
8.4~GHz, while the average X-ray count rate was 3550\ppm100~c/s.

One can decrease the noise level in the radio data by simply
increasing the integration time.  However, if the radio flux density
is strictly proportional to the X-ray flux, the signal-to-noise ratio
will decrease.  Using the radio data observed when the type~II burst
X-ray flux was $>$300~c/s yields a three-sigma upper limit of
255~\uJy\ on the radio emission during bursts while the average X-ray
type~II burst intensity was 2340\ppm120~c/s.  \label{sec:upplim}

\subsection{Relationship between X-ray and Radio Intensity}
\label{sub:xray-v-radio}

\begin{figure}[t!b]
\PSbound{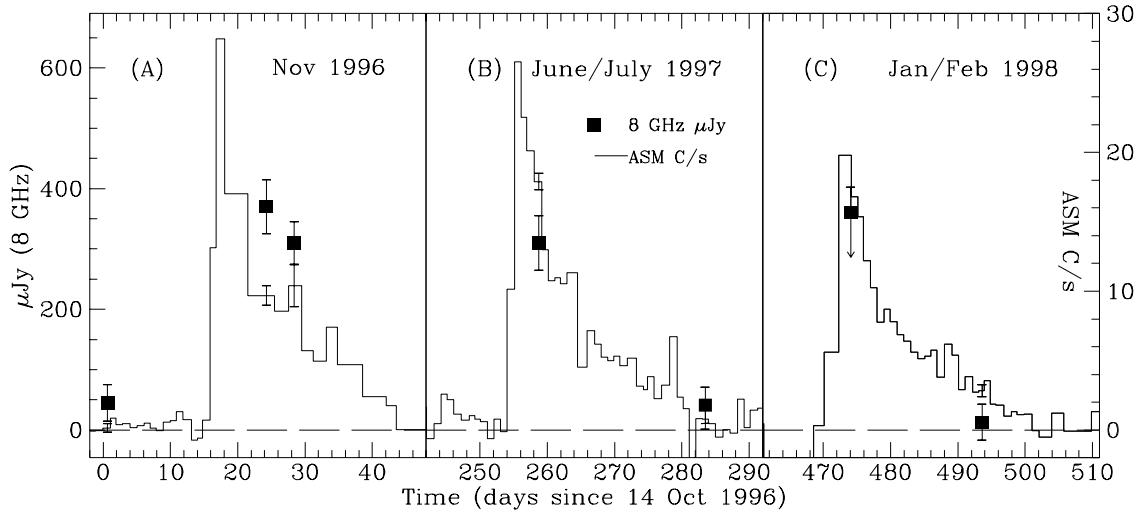}{15cm}{bbllx=40pt,bblly=445pt,bburx=580pt,bbury=695pt}
\caption{ \small \squeeze \label{fig:asmlc} 8.4~GHz flux density (VLA,
  and ATCA upper limit below 600 \ujy) of the proposed radio
  counterpart vs.~the \rxte/ASM (2-10 keV) daily-average count rate of
  the Rapid Burster as a function of time, near the November~1996
  (panel~A) and June/July~1997 (panel~B) and Jan/Feb~1998 (panel~C)
  outbursts.  The reference date is 14~October~1996, MJD~50370.  The
  scale of the radio flux density is on the left, the X-ray count rate
  is on the right.  The radio flux density turns ``on'' and ``off''
  with the 2-10~keV X-ray flux of the Rapid Burster. The probability
  of a randomly varying radio source duplicating the observed X-ray
  on/off cycle is 1--3\% (Section~\protect\ref{sec:sigXR}).  Radio
  upper limits are 3$\sigma$.  The ATCA data point for 1998~Feb~8 (day
  482) is off the figure with a $3\sigma$ upper limit of 930 \uJy.}
\end{figure}

In Figure~\ref{fig:asmlc}, we show the 8.4~GHz VLA flux densities
measured during three consecutive outbursts of the RB compared with
the \rxtez\ ASM X-ray intensity measurements.  
The radio outbursts and quiescent periods correspond well with the
X-ray state of the RB, indicating a possible relationship between the
radio and X-ray sources.

\begin{figure}[t!b]
\PSbound{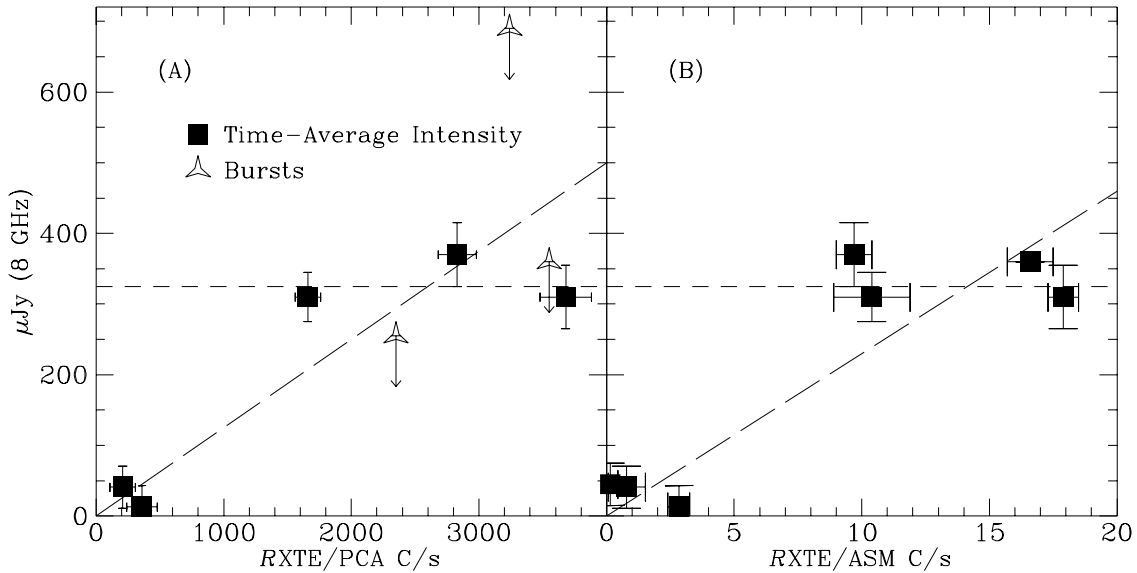}{15cm}{bbllx=36pt,bblly=417pt,bburx=570pt,bbury=690pt}
\caption{ \small \squeeze \label{fig:asmpca} The 8 GHz flux density
  vs. the simultaneous \rxte/PCA count rate (panel~A) and the
  contemporaneous \rxte/ASM count rate (panel~B).  The solid squares
  represent the time-averaged intensity (both X-ray bursts and
  persistent emission), while the open triangles $3\sigma$ upper
  limits derived from data taken only during type~II bursts.  (Note
  that the two most constraining of these points are derived from
  alternative analyses of the same data---see
  section~\ref{sec:Jan98}.)  In general, a linear correlation between
  X-ray and radio intensity does not provide a good fit to the data; a
  constant radio flux density at high time-averaged X-ray intensity is
  more consistent.  The radio flux densities include VLA (8.4 GHz) and
  ATCA (8.6 GHz) observations with sensitivity below 600 \uJy.  }
\end{figure}

For the ASM, the best-fit linear relation between the 8.4~GHz flux
density measured at the VLA and the ASM X-ray flux (forcing the line
through the origin) gives $S_{\rm8\,GHz} = 27\ppm1.7\,\mu{\rm
Jy}/({\rm 1~ASM~c/s})$.  However, this is a very poor fit, with a
reduced $\chi^2_\nu=5.6$ (for 6 degrees of freedom). For the PCA data,
the relation is $S_{\rm8\,GHz} = 125\ppm8\,\mu{\rm
Jy}/({\rm1000~PCA~c/s})$, again with a very poor fit (
$\chi^2_\nu=4.0$ for 4 degrees of freedom).  These data and the linear
fits are shown in Figure~\ref{fig:asmpca}.

It seems also plausible (observationally, if not theoretically) that
the radio source is ``on'' at a constant flux density (of 325\ppm25
\uJy\ at 8.4~GHz) whenever the X-ray source is above a threshold of
$\sim$3 ASM c/s ($\sim$500 PCA c/s).  In Figure~\ref{fig:asmpca} panel
(a), we also show the 3$\sigma$ upper limits for the excess radio
emission correlated with the average type~II burst intensity.  The two
most constraining upper limits (which were drawn from the same data;
see Section~\ref{sec:upplim}) are still marginally consistent with the
radio detections during periods of X-ray persistent emission of
comparable X-ray intensity.  The most constraining simultaneous radio
emission limit during the type~II bursts is discrepant at the
2.9$\sigma$ level.  

\section{Analysis of Radio Observations Using the Synchrotron Bubble Model}
\label{sec:results}

One possible description of our radio flux density measurements
involves a synchrotron bubble model \cite{vanderlaan66,hjellming95}.
In this model, the physical source of the radiation is a dense,
expanding bubble of plasma.  The relativistic electrons in the plasma
(assumed to have a more or less isotropic velocity distribution and a
power-law energy distribution) emit synchrotron radiation as their
trajectories are deflected by interactions with ambient or entangled
magnetic fields.  The resulting radio spectrum is strongly peaked with
a power law form at frequencies (much) below or above the peak
frequency.  This peak frequency decreases as the bubble expands in a
well-determined fashion: $\nu_m \sim\rho^{-(4\gamma+6)/ (\gamma+4)}$,
where $\nu_m$ is the peak frequency, $\rho$ is the normalized radius
of the bubble, and $\gamma$ is the power-law index of the electron
energy distribution, $N(E)\,dE\sim E^{-\gamma}\,dE$
\cite{vanderlaan66}.

A synchrotron bubble's time dependent radio emission is completely
determined if we set the time, $t_0(\nu_0)$, and the flux density at
spectral peak, $S_0(\nu_0)$, when the spectral peak reaches some
nominal frequency $\nu_0$ (which we will take to be the VLA X-band,
8.44~GHz).  Furthermore, one must define the functional form of the
bubble's radial expansion in time.  Here we adopt the convention of
previous authors \cite{vanderlaan66,hjellming95} and assume a
power-law form for this expansion: $\rho \sim t^{\slantfrac{1}{\alpha}}$.

The initial hypothesis that we wish to investigate is whether our
radio flux measurements may be explained by the evolution of a single
synchrotron bubble initiated at the start of each X-ray outburst.  We
perform this analysis for two primary reasons: first, models of this
nature, with time scales of the order of those observed in the radio
emission, have proven successful in modeling the radio outbursts of
other X-ray binaries \cite{hanthesis,hjellming95}; second, even if
the model is not successful, we expect it to give  insight into the
physical processes responsible for the radio emission we observe.

We determine the X-ray outburst start times by fitting the \rxtez\ ASM
light-curve to a functional form consisting of an onset time, a linear
rise in intensity, a peak intensity time, and an exponential decay --
similar to the form used by Guerriero \etal\ \cite*{bobg98}, except
that we do not include the secondary flares which are found by those
authors in two of the outbursts.  The results of our fits are
consistent with those of Guerriero \etal, and parameter uncertainties
are probably dominated by the systematics associated with this
simplified outburst intensity model.  Based on these fits, we
determine the time after outburst onset for each of the radio
observations.

\begin{figure}
\PSbox{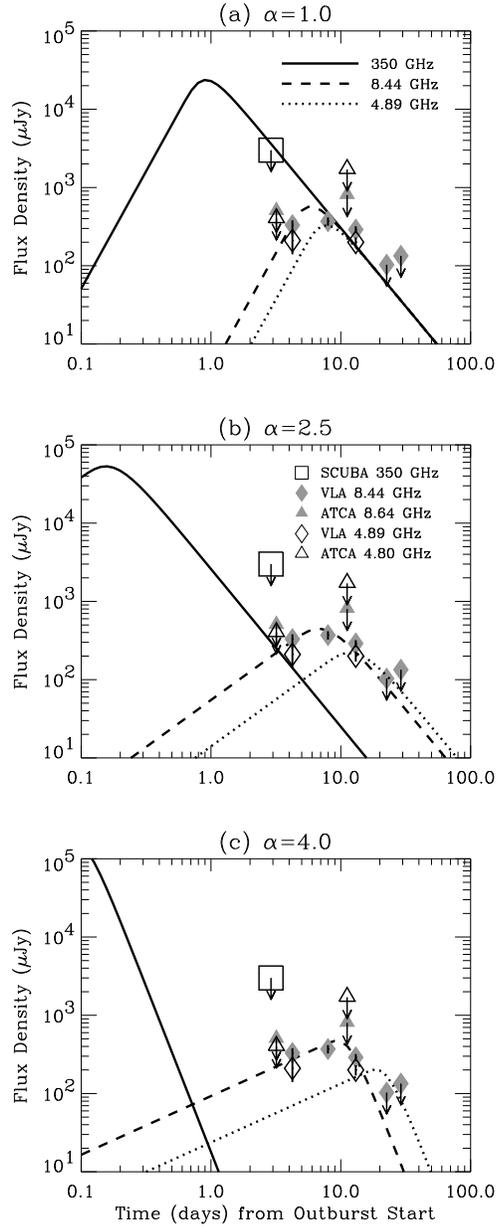}{7cm}
\caption{ \small \squeeze \label{fig:synall}
  Data and upper limits (3$\sigma$) from our radio observations, with
  best-fit synchrotron bubble model superposed for: {\bf (a)}
  $\alpha=1$ ($\chi^2_\nu=4.4$), {\bf (b)} $\alpha=2.5$
  ($\chi^2_\nu=2.2$), {\bf (c)} $\alpha=4$ ($\chi^2_\nu=1.9$).  Time
  is measured from the start of each Rapid Burster outburst, as
  determined by our model fits to the \rxte/ASM data (see
  Section~\protect\ref{sec:results}).  Lines indicate the predicted light
  curve of the synchrotron bubble event at three frequencies: 350~GHz
  (solid line), 8.44~GHz (dashed line) and 4.89~GHz (dotted line).
  The slightly different operational frequencies of the ATCA are
  effectively indistinguishable on the plot from the VLA frequencies
  shown.  Parameters of the model fits are given in
  Table~\protect\ref{tab:synch}.  Although this one-bubble model for
  the outbursts provides a reasonable fit to the observations, the
  derived parameters of the bubble's expansion are unphysical (see
  Section~\protect\ref{sub:bubble}).
} 
\end{figure}

We have performed fits to our radio observations using the synchrotron
bubble model of \citenp{vanderlaan66} Eqs.~11 \& 12; note however that his
Eq.\ 11 is incorrectly typeset and should read
\begin{equation}
        S(\nu,\rho) = S_{m0} (\nu/\nu_{m0})^{5/2} \rho^3
          \frac{ \left[ 1 - \exp\left( -\tau_m 
                        (\frac{\nu}{\nu_{m0}})^{-(\gamma+4)/2}
                        \rho^{-(2\gamma+3)} \right) \right] }
               { [ 1 - \exp(-\tau_m ) ] }
\end{equation}
where the variables are as defined in that work.  For purposes of this
analysis, we assume that the observed RB outbursts generate synchrotron
bubbles with identical physical properties and time evolution.  The
data are not sufficient to adequately constrain all four independent
parameters so we have fixed the expansion index $\alpha$ at three
canonical values: $\alpha=1$ corresponding to free expansion with
constant velocity, $\alpha=2.5$ corresponding to energy-conserving
expansion into an ambient medium (as in the Sedov phase of a supernova
remnant), and $\alpha=4$ resulting from a momentum-conserving (but not
adiabatic) expansion into an ambient medium \cite{hjellming95}.
Application of the geometric corrections to the van der Laan model
suggested by Hjellming and Johnston \cite*{hjellming88} altered the
fit parameters slightly but did not significantly improve the fits or
change the character of the solutions (the light curves changed by
less than 10\%). Therefore, we quote results from the simpler model.
The resulting synchrotron bubble parameters are given in
Table~\ref{tab:synch}
\begin{table}[t!b]
\footnotesize\squeeze
\begin{center}
\caption{\squeeze Synchrotron Bubble Model Fits to the Radio
     Observations\label{tab:synch}.  The models presented here had
     nine degrees of freedom. } 
\vspace{2mm}
\begin{tabular}{ccrrc}\tableline
 Expansion index & ~ & \multicolumn{2}{c}{Normalization (at 8.44 GHz)}    & Electron energy index \nl
 $\alpha$ (fixed) & Fit $\chi^2_\nu$ & \SzX, \uJy & \tzX,  days  & $\gamma$ \nl \tableline

 1 & 4.4 &  37.$^{+102}_{-3}$ & 27.9$^{+1.1}_{-13.7}$ & $<$1.07\tablenotemark{a} \nl
2.5& 2.3 & 342.$^{+35}_{-36}$ & 10.0$^{+1.5}_{-0.7}$  & 2.6$^{+4.4}_{-0.4}$ \nl
 4 & 1.7 & 438.$^{+61}_{-63}$ &  8.5$^{+4.9}_{-3.5}$  & 9.0$^{+5.0}_{-2.4}$ \nl
\tableline
\end{tabular}
\end{center}
{\small\squeeze $^{\it a}$Fixing $\alpha=1$ drove the fits to
           $\gamma=1$, the minimum physical value. $\gamma<1.07$ in
           this model with 90\% confidence.}
\end{table}
and the models are compared to the observations in Figure~\ref{fig:synall}.

As indicated in Table~\ref{tab:synch}, the models provide reasonable,
but statistically unacceptable, fits to the data ($\chi^2_\nu =$
1.7--4.4, for nine degrees of freedom).  Given our many simplifying
assumptions, particularly the assumption that all outbursts are
identical, this is perhaps not surprising.  We note that the free
expansion ($\alpha=1$) model fit is significantly worse than the other
two and violates the $3\sigma$ upper limit imposed by the
1998~January~30 SCUBA observation (Figure~\ref{fig:synall}a); for
these reasons we prefer the models which assume a deceleration of the
bubble's expansion by an ambient medium.

Even these models, however, are in the end physically unacceptable, as
may be determined by looking at the underlying physical parameters.
We can express $S_0$ in terms of the ambient magnetic field density,
$H_0$, and the angular extent of the source, $\theta_0$, at time \tzX,
as follows:
\begin{equation}
        S_0 = 0.85\, h(\gamma)\, \left(\frac{H_0}{1\,{\rm
        mG}}\right)^{-1/2}\, \left(\frac{\theta_0}{1\,{\rm
        mas}}\right)^{2}\,\, {\rm Jy},
\end{equation}
where $\gamma$ is the power-law index of the electron energy
distribution (2.6$^{+4.4}_{-0.4}$ for the adiabatic model and
9.0$^{+5.0}_{-2.4}$ for the momentum-conserving model), and
$h(\gamma)$ is a known function that varies from 4.1 to 1.04 as
$\gamma$ increases from 1 to 10:
\begin{equation}
        h(\gamma) = \frac{ \pi \Gamma(\frac{3\gamma+19}{12}) 
                        \Gamma(\frac{3\gamma-1}{12})
                        \Gamma(\frac{3}{4}) }
                  { \sqrt{6} \Gamma(\frac{3\gamma+2}{12})
                             \Gamma(\frac{3\gamma+22}{12})
                             \Gamma(\frac{5}{4}) },
\end{equation}
where $\Gamma$ is the Euler gamma function.  

Under the adiabatic assumption, the 8.44~GHz flux density of the
source when the spectral peak reaches that frequency is $\SzX =
340\ppm 35$~\uJy, whereas under the momentum-conserving assumption it
is 440\ppm 50~\uJy.  Thus the indications are that $\theta_{0} \sim
10^{-3}\,{\rm mas}$ at \tzX, under both models, for fields of the
order of a milligauss (with a weak \HmG{1/4}\ dependence on the actual
field strength).  At the distance of the RB (8.6~kpc) that corresponds
to a linear size of $d_0\sim 2$ light-seconds, and expansion
velocities at time \tzX\ of order 1~\kmsec, much too slow for the
expanding hot plasma that the model requires.  We are forced to
conclude that more complicated models are required to accurately
describe the observations (see Section~\ref{sub:bubble}).

\section{Evaluation of Counterpart Likelihood}
\label{sec:likelihood}

\subsection{Positional Coincidence}
\label{sec:sigpos}
First, we consider the probability that an unrelated radio source lies
close to the RB on the sky.  Micro-jansky source counts at 8.44~GHz by
Windhorst et al. \cite*{windhorst93} find a source density of
$1.14\times10^{-2}~{\rm arcmin^{-2}}$ for flux densities above
300~\uJy.  Therefore, the probability of an unrelated source brighter
than 300~\uJy\ falling within 8\farcs6 of the X-ray position of the RB
is $7\times10^{-4}$.  It should be noted that the \uJy\ sources are
typically associated with faint blue galaxies \cite{fomalont91} and
that the survey fields are not representative of the rich environment
of a globular cluster near the galactic center.  Because the
\uJy\ source population in an environments comparable to
Liller~1 is unknown, we do not use the probability computed above in
our evaluation of the counterpart likelihood which is based exclusively
on the correlation between the radio and X-ray lightcurves.

The best X-ray position for the RB \cite{grindlay84} places the radio
source 8\farcs6 away from the RB with $1\sigma$ X-ray and radio
uncertainties of 1\farcs6 and 0\farcs1, respectively.  Taken at face
value, this is significant at the 5.4$\sigma$ level, not including the
systematic uncertainty in the radio to optical (i.e.  \einsteinz\ star
tracker) reference frame shift.  This makes the radio source
identification with the RB seem less likely.  However, further
investigation reveals that the quoted Gaussian errors are an
inadequate description of the true error distribution.

The Einstein position for the Rapid Burster was determined from a
single pointing (four Einstein HRI pointings at Liller~1 were made,
but only one found the RB in outburst).  The 1\farcs6~$1\sigma$
uncertainty in this position is estimated from the dispersion of the
errors in the single-pointing Einstein positions of X-ray sources with
known optical counterparts \cite{grindlay84}.  An important
cross-check of the requisite star-tracker calibration (which dealt
with a number of complex systematic effects; \citenp{grindlay81}) was
provided by the Einstein globular cluster X-ray source program, which
performed multiple pointings at each of eight globular clusters
(GlCls) with known bright X-ray sources in order to determine the
positions of these sources to better than 1\farcs6 accuracy
\cite{grindlay81,grindlay84}.  These eight clusters were
NGC 104 (47 Tuc), NGC~1851, Terzan~2, Liller~1, NGC~6441, NGC~6624,
NGC~6712, and NGC~7078 (M15).

If the star-tracker calibration was successful in accounting for all
significant sources of systematic error, then we would expect the
resulting sample standard deviations, $s,$ in the positions of the
GlCl X-ray sources as derived from the multiple pointings at each, to
cluster strongly around $s = 1\farcs6$.  Examining the quoted
uncertainties of Grindlay et al.\ \cite*{grindlay84}, however, we find
that the actual $s$ values deviate from 1\farcs6 by up to a factor
of two --- $s = 0\farcs8$ for the X-ray source in NGC~6712 (5
pointings), while $s = 3\farcs2$ for the X-ray source in Terzan~2
(4 pointings).  A $\chi^2$ test shows that in NGC~6712, $s$ is too
small at the 98\% confidence level and that for Terzan~2, $s$ is too
large with 99.95\% confidence.  Together, these deviations are unlikely
to be statistical and they represent two out of the seven clusters
examined (recall that Liller~1 had only one pointing with a detection,
so its $s$ value is undetermined).

We conclude that there are probably remaining unaccounted-for
systematic effects in the Einstein aspect solutions.  These effects
average into the quoted 1\farcs6 error over many pointings, but
caused substantial non-Gaussian excursions within the context of the
GlCl X-ray source program.  Since the multiple pointings at each
cluster were typically executed over a time span of weeks to months,
the unaccounted-for systematics are not likely to be temporal in
nature, but rather to relate to position on the sky.  They may relate,
for example, to the density of suitable stars within the star-tracker
fields.  In this connection it is worth noting that Terzan~2 and
Liller~1 are the two sample GlCls nearest the Galactic Center, and are
only 3\arcdeg\ apart on the sky.

\subsection{Significance of the X-ray -- Radio Correlation}
\label{sec:sigXR}

The probability that the observed radio/X-ray behavior is produced by
an unrelated background radio source depends on the model for the
variability behavior of that source.  We consider first a source model
in which the background radio source varies randomly with a duty
cycle, $p$, and has a short auto-correlation time-scale allowing us to
consider each of our measurements to be statistically independent.

We take the radio source to be ``on'' or ``off'' when our 1$\sigma$
radio sensitivity is $<$70 \uJy\ and the source is or is not detected,
respectively, at the $3\sigma$ level.  Moreover, we define the Rapid
Burster to be ``on'' or ``off'' (for these purposes) when its one-day
average \rxtez\ ASM count rate is greater or less than 4 cts/sec (see
Figure~\ref{fig:asmpca} and Section~\ref{sub:xray-v-radio}).  Based on
these assumptions, the probability of an unrelated variable background
radio source mimicking the X-ray on/off state of the RB is the product
of the normalized probability distribution for obtaining the number of
radio ``on'' and ``off'' observations for a given $p$, and the
probability distribution of observing the radio and X-ray sources to
be ``on'' and ``off'' simultaneously for a given $p$, integrated over
all values of $p$:

\begin{equation}
P = \frac{\int_0^1 p^N (1-p)^M \, p^{N_s} (1-p)^{M_s} \, dp}
            {\int_0^1 p^N (1-p)^M \, dp},
\end{equation}

\noindent where $N$ is the total number of times the radio source is
observed ``on'', $M$ is the total number of times the radio source is
observed ``off'', $N_s$ is the number of times the radio source and
X-ray source are observed on simultaneously, and $M_s$ is the number
of times the radio source and X-ray source are observed off
simultaneously. We have assumed a uniform prior distribution for $p$
itself.  For our observations (\cf~Table~\ref{tab:radio}), $N=3$,
$M=3$, $N_s=3$, $M_s=3$, which yields a probability of 1.2\% for an
unrelated radio source mimicking the observed on-off behavior of the
RB.  This indicates that it is unlikely that the observed radio/X-ray
correlation is produced by a randomly varying background radio source.

More complex models for the radio source might postulate that it
occasionally turns ``on'' and remains so for $T_{\rm on}$ days
(auto-correlation time scale).  We can estimate the probability of a
chance correlation of such a source with the Rapid Burster, without
performing involved Monte Carlo simulations, by making use of the
framework developed above.  If we speculate that $T_{\rm on}$ is such
that the two detections during Nov~1996 (separated by 5~days) are
perfectly correlated and all other observations are uncorrelated, then
we have 5 coincident X-ray/radio observations ($N=N_s=2$, $M=M_s=3$),
and the probability of a chance correlation is 2.6\%.  Note that if
the auto-correlation time scale is shorter than 5 days then all six
observations are statistically independent as above.  Note also that
the auto-correlation time scale cannot be much longer than 5 days,
because as it approaches $\sim$20 days it approaches the observed
radio on/off time (that is, the initial non-detection followed by a
detection during Nov~1996, and the detection followed by a
non-detection in June/July~1997).

One might reasonably be concerned about the possibility of an
unrelated radio source that turns ``on'' suddenly and remains so for
an extended period of time.  The radio data do not exclude the
possibility that such a source was present and remained ``on'' during
the unsampled period between the Nov~1996 and June/July~1997 outbursts
of the RB.  In this case, we have only three independent measurements
of coincidence with the RB.  However, the radio source is observed to
make the transition from ``off'' to ``on'' (and the reverse) in
\approxlt 25~d.  The radio data show one ``turn-on'' and one
``turn-off'' both of which are well sampled and coincident with the
X-ray behavior of the RB.  If an unrelated radio source turns ``on''
every $\sim$300~d, then the probability of the ``turn-on'' coincidence
observed in Nov~1996 is $\sim$8\%.  The slow decay of the RB X-ray
outburst makes a ``turn-off'' coincidence somewhat more probable.  If
we cannot distinguish ``turn-offs'' that are separated by $\pm 10$~d,
then the probability of the observed coincidence in the June/July~1997
outburst is $\sim$15\%.  If the two events are independent, the
probability of this type of source mimicking the observed RB X-ray
turn-on/off behavior is $\sim$1\%.

The population of \uJy\ radio transients and their outburst properties
are not well studied.  The probability of an unrelated, variable radio
source having the time-dependent characteristics above and lying this
close to the RB is somewhat less than unity although we have neglected
this in the above calculations.  It seems reasonable to assume the
most conservative of the above approaches as the upper limit; we
therefore find that the upper limit on the probability of a flaring
background radio source mimicking the X-ray behavior of the RB is 3\%.

\subsection{Possible Relationship Between the \fg\ Radio Object and the
Present Object}

The relationship between the variable radio source we observe and the
steep spectrum source ($\alpha = -2$) observed by \fg\ is not clear.
To within the astrometric uncertainty (1\farcs5, 3$\sigma$), the two
sources are at the same position on the sky.  If \fg's interpretation
of their source is correct and it represents the integrated emission
of a population of radio pulsars, then it seems likely that their
radio position is the center of the GC.  (The $\sim$2\arcsec\
separation between the \fg\ radio source and the optical center of
Liller~1 is consistent with the 1\arcsec\ uncertainty in the absolute
optical astrometry; see Figure~\ref{fig:rbradio}). For a 6\farcs5
core-radius GC \cite{kleinmann76,picard95}, the 1\farcs5 radio
error circle represents 12\% of the optical light of the GC.  Thus, it
is not unlikely that (if the RB is associated with the GC) the radio
counterpart to the Rapid Burster would also lie close to the center of
the GC.

The three flux-density measurements of the \fg\ source were made over
three different (sometimes overlapping) epochs.  Combining the three
measurements into a single spectrum (which was found to be steep)
assumes that the source is not variable.  The conflicting 1.5~GHz
measurements of \fg\ and Johnston \etal\ may indicate variability of a
factor of two or more (2$\sigma$) between April 1990 and May 1993.
The X-ray state of the Rapid Burster during these observations is
unknown, and it may be that it was X-ray active during \fg's 1.5~GHz
observation, providing the additional radio flux above that expected
for an underlying population of radio pulsars in Liller~1.  The
relationship between the \fg\ radio source and the RB could be
illuminated by radio measurements of the same epoch at 0.33, 1.5 and
4.5~GHz, taken both while the RB is in X-ray outburst and in quiescence.

\subsection{Conclusion -- A Likely Radio Counterpart}

The probability that a serendipitously located variable radio source
would mimic the Rapid Burster X-ray state as has been observed is
small (1--3\%), but not dismissably so. The number and distribution of
faint, variable radio sources toward the Galactic Center is not well
known. There has been at least one other recent instance where a radio
source, variable on a time-scale of $\sim$days, was discovered
$\sim3^\prime$ from a bright X-ray source (although the X-ray flux was not
correlated; \citenp{frail96a,frail96b}).  In addition, globular
clusters are known to harbor both millisecond radio pulsars and
accreting X-ray sources, so the appearance of an unrelated variable
radio source in Liller~1 must be considered more likely than for a
random field.  Other than the correlated radio and X-ray states and
the marginal consistency of the radio spectrum with synchrotron-bubble
models, we have not observed any radio behavior which would tie this
object uniquely to the Rapid Burster.

The apparent positional discrepancy between the radio source and RB is
likely explained by the non-Gaussian distribution of the X-ray position
error as discussed above and acknowledged by Grindlay et al.\
\cite*{grindlay84} and Grindlay \cite*{grindlay98}.  Proper motion of
the RB will probably contribute only negligibly to the discrepancy,
given the association with Liller~1, even though an interval of 20
years separates the \einsteinz\ and radio observations (a proper
motion of $\sim$0\farcs5 per year is required).  Improvement upon
the X-ray localization will be obtained during an approved \axafz\
observation, which should determine the X-ray position to \approxlt
0\farcs5, thus confirming or excluding this radio counterpart.

At present, we identify this radio source as a likely radio
counterpart of the Rapid Burster.  Apart from the \axafz\ observation,
the case could be strengthened by observing bursts from this location
(either in radio or IR), by continued observations in X-ray and radio
of correlated on/off behavior, by observation of larger swings in the
X-ray and radio fluxes which would permit a definite X-ray/radio
correlation to emerge, or by discovering other radio behavior which is
correlated with X-ray behavior of the Rapid Burster (such as short
time-scale variability).  For example, a program of 12 short VLA
observations reaching a noise level of 45 \uJy/beam taken at two-week
intervals and correlated with contemporaneous \rxtez\ ASM observations
would reduce the probability of an unrelated source mimicking the RB
to 0.03\%, even if the RB remains quiescent in X-rays and there are no
radio detections.

\section{Discussion of the Radio Observations}
\label{sec:discuss}

The radio observations of the proposed radio counterpart on
1996~November~11.88 determine that the radio spectral slope is flat to
inverted with $\alpha$=0.9\ppm0.3 ($S_\nu \propto \nu^\alpha$).

The JCMT/SCUBA observation at 350~GHz is, to date, the
earliest radio observation relative to the beginning of an X-ray
outburst.  A few hours prior to this observation, a PCA observation
measured the RB PE to be 4000\ppm100 c/s.  Using the measured radio
spectral slope ($\alpha$=0.9\ppm0.3) and the PCA/radio conversion at
8.4~GHz, the extrapolated radio spectrum gives a 350~GHz flux density of
4.5--41~mJy (with uncertainty dominated by the spectral slope).  This
is well above the 3 mJy $3\sigma$ upper limit obtained at the JCMT
indicating that the radio emission is not a simple power-law spectrum,
proportional in intensity to the instantaneous X-ray intensity.  This
non-detection is also the crucial observation in ruling out one class
of synchrotron bubble models (see Section~\ref{sec:results}).

\label{sub:bubble}

% Note -- Hjellming and Han have a typo: GS2000+35=GS2000+25

Synchrotron bubble behavior, associated with the outburst of an X-ray
transient, has been observed on many occasions; Hjellming \& Han
\cite*{hjellming95} show radio data and fits for A0620$-$00, Cen~X-4,
GS~2000+25, Aql~X-1, GS~2023+338 (V404~Cyg), and GRS~1124$-$683 (see
also \citenp{hanthesis}).  In that respect the detection of radio
emission from the Rapid Burster during its outbursts is not
particularly surprising.

The physical parameters derived from our model fits, however, indicate
expansion velocities of $\sim$1~\kmsec\ about 10~d after the outburst
start which is well below the physical lower limit set by the sound
speed of the hot plasma, $c_{\rm s} \sim 0.1 \sqrt{T_{\rm K}}$ \kmsec\
(where $T_{\rm K}$ is the temperature of the plasma in Kelvin,
$\sim$10$^7$ in SS~433 -- \citenp{hjellming88}).

Our assumption of a generic bubble event accompanying each outburst is
therefore likely to be flawed and we must consider more complex
models.  For example, radio emission over the course of an outburst
may result from the summed emission of a succession of synchrotron
bubbles, each of which expands at high speed and therefore brightens
and fades (at a given frequency) much more quickly than the emission
at large.  If the flux levels that we see are produced by $T_{\rm
K}\sim 10^7$ plasma, then the expected rise times for the synchrotron
bubbles are $\sim$1 hour.  Alternatively, each of our radio detections
may simply have caught a single fast bubble in the midst of its
expansion; in that case, the timing of our detections would indicate
how the rate of these bubble ejections changes over the course of an
outburst.  Relevant to both of these possibilities, it is worth noting
that the radio flares from GRS~1915+105 have rise and decay times of
$\sim$1 hour \cite{mirabel98,fender98}.

We are therefore motivated to consider whether the type~II X-ray
bursts of the Rapid Burster might produce individual synchrotron
bubbles, with associated infrared and radio emission.  The synchrotron
bubbles of GRS~1915+105 have been shown to be related to active
accretion, as determined by simultaneous X-ray observations
\cite{mirabel98,eik98a}.  Scaling down the brightness of the infrared
emission seen in that source by the factor of ten difference in X-ray
flux between it and the RB suggests that there may be mJy infrared flares
($K \approx 14$) during RB type~II X-ray bursts, if synchrotron
bubbles are indeed being formed.  We are currently pursuing short
time-resolution IR observations of the RB during its next outburst to
test this hypothesis; naturally, observation of any bursting
counterpart to the RB will confirm or reject the VLA counterpart
proposed here.

The radio detections we report here could not themselves have been
produced by the type~II X-ray bursts; the radio detections were made
prior to day 12 of each outburst, while the type~II X-ray bursts were
not observed until after day 14 in every case.

The observations of 1997~July~24 were carried out when the VLA was in
its new CS (C-short) configuration.  The combination of the array
configuration with low-elevation observing yielded many short
projected baselines with lengths all the way down to the antenna
separation (25~m).  Heavily tapered maps of these data reveal an
extended source with peak flux density of approximately 3~mJy (at
8.44~GHz) that is at least as large as the primary beam.  The axis of
symmetry of the extended source lies at its closest point
approximately $1^\prime$ north-east from the RB, which corresponds to
2.5~pc at the 8.6~kpc distance of the RB.  The source is probably
unrelated to the RB but its presence is an important consideration
for future observations that include short interferometric baselines.

\section{Conclusions}
\label{sec:conclusions}

We have detected a likely radio counterpart for the Rapid Burster,
with radio emission correlated with the X-ray outbursts.  The
likelihood of unrelated variable radio source duplicating the observed
correlation between the X-ray flux and radio flux density is low
(1--3\%), but not dismissably so.  There is an apparent discrepancy
between the X-ray position of the RB and the radio counterpart but it
is likely due to the non-Gaussian distribution of the X-ray position
errors.  Confirmation of the counterpart from additional observations
-- an already approved \axafz\ observation, further radio observations
while the \rxtez\ ASM is still operational, or possibly infrared
observations while the RB is in outburst -- is required.  The time and
spectral evolution of the radio source, while not physically
interpretable as a full-outburst synchrotron-bubble (as seen in some
other transient X-ray binaries), may be due to $\sim$hour-long radio
flares such as have been seen from the superluminal-jet source
GRS~1915+105.

Our lower-limit on the time delay between X-ray emission and radio
emission from the type~II bursts (\approxgt 1 sec) is consistent with
the delay expected from a synchrotron bubble.  Our observation of a
persistent radio source $\sim$5 days after the start of active
accretion (i.e.\ an outburst) sets an upper limit for the radio vs.\
X-ray time delay.  The correlation of radio flux density with
persistent X-ray intensity in this system indicates that the radio
flux density is related to active accretion onto the surface of the
neutron star -- as accretion is responsible for the X-ray outburst.
This suggests that excess radio emission may be produced during the
type~II X-ray bursts, which are themselves accretion driven.  However,
our observations produce no evidence of simultaneous radio/X-ray
bursts, marginally constraining them (2.9$\sigma$) to be below the
level that we observe during periods of comparable X-ray flux in
persistent emission.  This may imply that radio bursts at 8.4~GHz do
not occur simultaneously with X-ray type~II bursts.

\section{Acknowledgements}
\label{sec:acknowledgements}
%\acknowledgements

We are grateful to Evan Smith, Jean Swank, and the XTE Science
Operations Facility staff for their efficient processing of the
several XTE TOO observations covered in this work.  We are also
indebted to VLA observers and staff Phillip Hicks, Robert Hjellming,
Rick Perley, Michael Rupen, and Ken Sowinski, who graciously gave us
time to perform observations of the RB, to Barry Clark who worked on
short-notice to help find the time to make observations during
critical periods, to Tasso Tzioumis who performed the observations at
ATCA, and to Ian Robson who made possible the observations at the
JCMT.  We are grateful to M.~van~der~Klis and the anonymous referee
for their detailed and helpful comments on the manuscript.  This work
has been supported under NASA Grant NAG5-7481.  JvP acknowledges the
support of NASA under grant NAG5-7414.  RPF was supported during the
period of this research initially by ASTRON grant 781-76-017 and
subsequently by the EC Marie Curie Fellowship ERBFMBICT 972436. CBM
thanks the University of Groningen for its support in the form of a
Kapteyn Institute Postdoctoral Fellowship. RER thanks his host
J.~Tr\"umper of Max-Planck-Institut f\"ur Extreterrestrische Physik,
where this work began, and his host Lars~Bildsten of UC~Berkeley,
where this work was completed.

\end{document}